\documentclass[12pt]{article}
\usepackage{epsf}
\usepackage{amsmath}
\usepackage{graphics}
\usepackage{cite}

\renewcommand{\thefootnote}{\#\arabic{footnote}}
\begin{document}

\newcommand{\gtrsim}{ \mathop{}_{\textstyle \sim}^{\textstyle >} }
\newcommand{\lesssim}{ \mathop{}_{\textstyle \sim}^{\textstyle <} }

\newcommand{\rem}[1]{{\bf #1}}

\renewcommand{\thefootnote}{\fnsymbol{footnote}}
\setcounter{footnote}{0}
\begin{titlepage}

\def\thefootnote{\fnsymbol{footnote}}

\hfill arXiv:08mm.nnnn [hep-ph]

\begin{center}

\vskip .5in

\bigskip
\bigskip
{\Large \bf Strong Incompatability of Tribimaximal Mixing
with Quark-Lepton Complementarity}

\vskip .45in

{\bf Paul H. Frampton\footnote{frampton@physics.unc.edu}
and Shinya Matsuzaki\footnote{synya@physics.unc.edu}}

\bigskip
\bigskip

{Department of Physics and Astronomy, UNC-Chapel Hill, NC 27599.}

\end{center}

\vskip .4in

\begin{abstract}
Although tribimaximal mixing (TBM) and
quark-lepton complementarity (QLC)
are both compatible with experimental data
within one standard deviation ($ 1 \sigma$),
the TBM and QLC assumptions are mutually
exclusive by $47 \sigma$.
\end{abstract}

\end{titlepage}

\renewcommand{\thepage}{\arabic{page}}
\setcounter{page}{1}
\renewcommand{\thefootnote}{\#\arabic{footnote}}

\newpage

Two popular assumptions in particle phenomenology
model building are tribimaximal mixing (TBM)
and quark-lepton complementarity(QLC). 
In this article we wish to point out that
because both TBM and QLC involve exact angles
it is dangerous to 
attempt to combine TBM with QLC.
To illustrate our point, it suffices to write down
some relationships involving experimentally determined
values of mixing angles and exact angles
predicted on the basis of different theoretical assumptions.

\bigskip
\bigskip

In TBM the assumption \cite{HPS} is that the neutrino
mixing angle $\theta_{12}$ satisfies
\begin{equation}
\tan \theta_{12} = \frac{1}{\sqrt{2}}
\label{TBM}
\end{equation}
corresponding to $\theta_{12} = 35.26...^o$.
The experimental value\cite{PDG2008} 
\begin{equation}
(\theta_{12})^{Experiment}  = 33.9^o \pm 1.4^o
\label{nuExpt}
\end{equation}
agrees within 
one standard deviation ($1 \sigma$).

\bigskip
\bigskip

In QLC the assumption is\cite{Raidal,Smirnov,FMoh}
that $\theta_{12}$ and the quark mixing angle $\Theta_{12}$ satisfy exactly 
\begin{equation}
\theta_{12} + \Theta_{12} = 45.00...^o
\label{QLC}
\end{equation}
and the experimental value\cite{PDG2008}
\begin{equation}
(\Theta_{12})^{Experiment} = 13.05 \pm 0.07^o.
\label{quarkExpt}
\end{equation}
together with Eq.(\ref{nuExpt})
agrees within $1 \sigma$.

\bigskip
\bigskip

If, however, we assume both TBM and QLC we predict exactly
\begin{equation}
\left( \Theta_{12} \right)^{TBM+QLC} = 9.74...^o
\label{TBMQLC}
\end{equation}
which disagrees with experiment, Eq.(\ref{quarkExpt}),
by an astonishing $47 \sigma$. Thus, any 
model which predicts TBM and QLC
must conflict with experiment
\footnote{It is worth noting that in some models ({\it e.g.} \cite{Altarelli})
higher order irrelevant operators do significantly modify predictions
for $\theta_{12}$ and could therefore possibly reconcile seemingly
conflicting assumptions made at the lowest order.}

\bigskip
\bigskip

Another theoretical assumption introduces a further exact angle.
From the viewpoint of $T^{'}$ flavor symmetry where $T^{'}$ 
is the binary tetrahedral group, Eq.(\ref{QLC})
is replaced by a different
relationship between $\theta_{12}$ and $\Theta_{12}$
involving a third exact angle\cite{Tprime}
\begin{equation}
\left(  \theta_{12}   + \Theta_{12} \right)^{T'}   = 
\left( \tan^{-1} (1/\sqrt{2}) \right)
+ \left( \frac{1}{2} \tan^{-1} (\frac{1}{3} \sqrt{2} ) \right)
= 47.88...^{o}
\label{TPRIME}
\end{equation}
so within the context of $T^{'}$ flavor symmetry there appears no
support for
the assumption of 
the exact angle $45.00...^o$ used in quark-lepton complementarity, Eq.(\ref{QLC}).

\bigskip

Above, we have denoted exact angles 
by three periods of ellipsis ... 
to emphasize the infinite accuracy of
prediction and strong incompatibility between two of them.
The emergence of exact angle predictions for flavor mixing
angles is an interesting development where a dot-dot-dot 
ellipsis signifies an unknown combination of theoretical correction and 
deviation from experiment.
The most important experimental data are 
those on $\theta_{12}$ to illuminate which
exact angle mentioned in this article 
merits further study.

\vspace{5.5in}

\begin{center}
{\bf Acknowledgement}
\end{center}

\noindent This work was supported by U.S. Department of Energy grant number
DE-FG02-06ER41418.

\end{document}